# Control of protein activity by photoinduced spin polarized charge reorganization

Shirsendu Ghosh[1,#,&], Koyel Banerjee-Ghosh[1,2,#], Dorit Levy[1], David Scheerer[1], Inbal Riven[1], Jieun Shin[3], Harry B. Gray[3], Ron Naaman[1,*] and Gilad Haran[1,*]

[1]Department of Chemical and Biological Physics, Weizmann Institute of Science, Rehovot, Israel

[2]Department of Chemistry, Indian Institute of Technology Hyderabad, Kandi, Sangareddy-502285, Telangana, India.

[3]Beckman Institute, California Institute of Technology, Pasadena, California 91125, United States

*Corresponding authors: Gilad Haran (email: gilad.haran@weizmann.ac.il) and Ron Naaman (email: ron.naaman@weizmann.ac.il).

[#]S.G. and K.B.-G. contributed equally to this work.

[&]S.G. current address: Department of Chemistry and Chemical Biology, Cornell University, Ithaca.

**Abstract**

Considerable electric fields are present within living cells, and the role of bioelectricity has been well established at the organismal level. Yet little is known about electric-field effects on protein function. Here we use phototriggered charge injection from a site-specifically attached ruthenium photosensitizer to directly demonstrate the effects of charge redistribution within a protein. We find that binding of an antibody to phosphoglycerate kinase (PGK) is increased two folds under illumination. Remarkably, illumination is found to suppress the enzymatic activity of PGK by a factor as large as three. These responses are sensitive to the photosensitizer position on the protein. Surprisingly, left (but not right) circularly polarized light elicits these responses, indicating that the electrons involved in the observed dynamics are spin polarized, due to spin filtration by protein chiral structures. Our results directly establish the contribution of electrical polarization as an allosteric signal within proteins. Future experiments with phototriggered charge injection will allow delineation of charge rearrangement pathways within proteins and will further depict their effects on protein function.

**Significance Statement**

The role of well-placed charges within proteins in mediating biological functions, from protein-protein association to enzyme kinetics, is well documented. Here we go beyond this static picture and show that charge motions can exert significant effects on protein function. Injecting charge from a photosensitizer, we demonstrate a three-fold decrease in enzymatic activity and a two-fold increase of antibody-antigen binding. These effects depend on the specific position of the photosensitizer on the protein. Our results point to charge reorganization as a form of allostery that complements known allosteric mechanisms such as conformational changes and dynamics.

\body

**Introduction**

Biomolecules within the living cell are subject to extensive electrical fields, particularly next to membranes(1). Indeed, a role for bioelectricity has been well established at the organismal level (2). While the importance of electrostatics in protein functions such as protein-protein association and enzymatic activity has been well documented (3), very little is known on how biomolecules respond to external electric fields, or in other words, what may be the potential contribution of polarizability to protein function. Multiple protein activities involve electrostatic effects (3). For example, it is recognized that the association kinetics of proteins can be accelerated by charged residues positioned close to the interaction sites on their surfaces (4). Recent work on enzyme catalysis has given rise to a picture of pre-organized charges at catalytic sites, directly influencing substrate molecules and lowering enzymatic reaction barriers in this manner (3, 5, 6). These mechanisms for charge influence on protein function invoke essentially fixed charge distributions, and do not take into account the potential role of charge regulation and reorganization due to external electric fields(7). Yet, it is important to appreciate that any interaction between two proteins, as well as between a protein and other molecular species, involves the formation of an effective electric field that results from the difference in electrochemical potentials of the two interacting bodies.

Since proteins have low dielectric interiors, variations in charge positions and electric fields that result from interactions, may have relatively long-range effects. Due to their internal conformational dynamics, as well as the presence of titratable side chains, proteins may possess significant polarizability values. Recent simulations from Takano and coworkers (8, 9) and experimental work from our labs (10, 11) have indeed hinted at a role for charge reorganization as an allosteric signal in proteins. Here we decisively establish this role by studying the effect of phototriggered charge injection on both protein-protein association kinetics and enzyme kinetics. We find a rich spectrum of responses that depends on the position of the photoexcited group as well as on the spin polarization of the rearranging charges. The spin dependence is likely associated with the chiral induced spin selectivity effect (12).

**Results**

**Modulating protein-protein association**

We site-specifically labeled phosphoglycerate kinase (PGK), a 415-residue protein, with the photosensitizer ([Ru(2,2′-bipyridine)2(5-iodoacetamido 1,10-phenanthroline)]$^{2+}$ (Ru) (13). In particular, we created the mutant C97S/Q9C, in which the native cysteine at position 97 was changed to a serine, and a cysteine residue was inserted at position 9 (Fig. 1). Ru can inject either an electron or a hole into the protein, potentially modulating the charge distribution (i.e., the electric polarization) within the protein. We first studied the binding of an anti-His antibody to a polyhistidine tag at the C terminus of PGK (Fig 2A). The Ru-PGK construct was attached to a gold surface to facilitate uniform illumination and readout of antibody-antigen interaction. The antibody molecules were labeled with the dye Alexa 647, which allowed counting individual events of protein-protein association at the surface at different times, following the addition of the antibody to the solution. The experiment was performed either under illumination with a linearly polarized (LP) 470 nm laser or in the dark (Fig 2B-E). The kinetics traces in Fig. 2F demonstrate that under illumination association was significantly enhanced at early times. In particular, at 2 s, illumination increased the association rate by a factor of 2.25±0.05. At longer times the difference between the two sets decreased, reaching a similar value at 8 s, due to saturation of the binding of antibody molecules to the surface. In the rest of the article, we will therefore report only rate differences at 2 s. The experiment was repeated with PGK molecules that were not labeled with Ru, and no effect of illumination was observed (Fig. 2G). We further repeated the same experiment on glass to rule out any potential contribution of the gold surface, and the results were similar (Fig. 2H).

As it is known that electron transport through a protein may be spin selective, due to the chirality of the protein and its secondary structure (12, 14, 15), we asked whether illumination with circularly polarized light can modulate the observed effect. The experiment on the gold surface was therefore repeated with either right or left circularly polarized light. Circularly polarized light is likely to generate excitations with one spin state(16), so that the injected charge into the protein (either positive or negative) would be spin polarized. Remarkably, the enhancement of the association kinetics was observed only with left circularly polarized (LCP) light, and not with right circularly polarized (RCP) light (Fig. 2I). These results indicate, within the experimental uncertainty, that the whole photoinduced effect an outcome of essentially a single spin polarization, suggesting in turn that the charge reorganization within the protein is spin selective.

To test the position dependence of the charge reorganization effect on association kinetics, the Ru complex was moved to residue 290, using the mutant C97S/S290C (Fig. 1). At this position, the photosensitizer is much further away from the His-tag at the C terminus compared to the previous position; the distance from residue 290 to the C terminus, residue 415, is 55 Å, based on the crystal structure 3PGK, while from residue 9 it is only ~10 Å. Repeating the same experiment, it was found that illumination (either LP, LCP or RCP) had only a minor effect on the association reaction (Fig. 2J), pointing to a significant position dependence of the effect.

**Controlling enzymatic activity**

We then turned to measure the effect of photosensitization on the catalytic reaction of PGK. The enzyme catalyzes the transfer of a phosphate group from ATP to 3-phosphoglycerate (3-PG), producing ADP and 1,3-bisphosphoglycerate (1,3-BPG) (Figure 3A). To observe a robust reaction on a surface, the His-tag at the C terminus of PGK was used to attach protein molecules to a supported lipid bilayer formed on a glass substrate (Fig. 3A). The turnover of surface-bound enzyme molecules was measured using a coupled assay, and the kinetics were gauged through a change in NADH absorbance (17). Based on the slopes of the kinetics curves in Fig. 3B-E (Supporting Table 2), and assuming a surface density of PGK molecules of $\sim 5 \cdot 10^{11}/cm^2$ (somewhat lower than expected for a close-packed layer of the protein), we calculated a turnover rate of ~200 $s^{-1}$ for Q9C PGK and S290C PGK in the dark. This turnover rate is quite close to the value measured in solution with C97S PGK (226.9±7.3 $s^{-1}$).

Remarkably, with Ru at position 290, the enzymatic rate decreased under illumination by a factor of 3.3±0.2 (Fig. 3B). As above, this rate reduction was induced by either LP or LCP illumination, but not under RCP illumination. The effect could be observed in a single experiment: when light was turned off, the slope of the kinetics curve increased (Figure 3C). In the absence of Ru, no effect of illumination was observed (Fig. 3C and Supporting Fig. 1A). When the Ru complex was moved to position 9, an illumination effect was still observed, but it was significantly reduced to a factor of only 1.8±0.1 (Fig 3D); as above, the slope increased when light was turned off, and no illumination effect was observed in the absence of Ru (Fig. 3E and Supporting Fig. 1B).

**Discussion**

When a protein interacts with a charged molecule/protein, charge rearrangement occurs within the protein, which may affect the interaction between the protein and the other species. The extent of charge rearrangement depends on the polarizability of the protein, and therefore polarizability may affect both interaction between proteins and enzymatic activity. Upon excitation of a photosensitizer, charge can be injected into the protein, hence affecting its polarizability, thereby modulating the effect discussed above. Charge injection can involve either an electron or a hole, and might potentially be only partial, leading in either way to an effect on the charge distribution within the protein. However, since the protein is chiral, any charge injection would be spin dependent due to the chiral induced spin selective (CISS) effect, as shown by Naaman and coworkers in multiple studies (12). Exciting the dye with circularly polarized light causes one spin to be preferentially excited. Due to the CISS effect, one specific spin can be injected more efficiently into the protein. Therefore, one circular polarization is more effective than the other. Excitation with the 'correct' circular polarization would lead to charge injection into the protein and to a charge-separated species that would typically have a much longer lifetime than the usual excitation lifetime of the molecule. On the other hand, excitation of the 'wrong' circular polarization would not lead to charge injection, and the excited state would relax quickly, either radiatively or non-radiatively.

Our results indeed indicate a significant effect of charge injection from the photosensitizer Ru into the protein both on association with an antibody and on its enzymatic reaction. Notably, the effects we measure depend on the polarization of light and in particular, respond to only one circular polarization. Our findings strongly support the notion that charge reorganization is involved, as it has been established that the motion of charge through a chiral potential is spin selective and is affected by the protein secondary structure (15). This spin dependence is due to the chiral-induced spin selectivity effect (12). Importantly, it has been shown previously that spin polarization enables long range charge transfer through chiral systems (18).

Specifically for our systems, we can only speculate on the exact effect of charge injection and in which direction charge is transferred. In the case of the antibody-protein interaction, since the antibody is directed to the His-tag, we observe the C-terminal region of the protein (where the His-tag is connected) and find that it is in general more negative. Clearly, the protein-protein association reaction would benefit from this region being even more negative, meaning that an electron would likely be injected from the photosensitizer. In the case of the enzymatic reaction, charge reorganization may affect substrate binding by making the active site more negatively charged and changing its interaction with the negatively charged substrate molecules. Additionally, charge reorganization may affect the catalytic mechanism itself. We cannot be more specific about this aspect at this moment of time. In any case, since charge reorganization is found to be sensitive to circularly polarized light, it is likely that $\alpha$-helical structures of the protein are involved, as $\alpha$-helices have been implicated as good spin filters (15).

The photoinduced charge injection effect we observe here depends on the distance from the active site involved, rather than on the sequence separation. Thus, for protein-protein association at the C terminus of PGK, Ru at position 9 had a strong effect, while Ru at position 290 had no effect. A similar picture arose also for the enzymatic activity of PGK, though now Ru at position 290 (close to the ATP binding site) showed double the effect of Ru at position 9.

The findings here, combined with previous studies (10, 11), point to a new role of charge reorganization, or of polarizability, in modulating protein activities. Surprisingly, not much is known about the involvement of polarizability in protein function, though the development of polarizable force fields for molecular dynamics simulations of biomolecules in recent years may change this situation(19). The role of charged protein residues in enzymatic catalysis has been discussed extensively by Warshel and coworkers(3), who emphasized the contribution of charges that are pre-organized to reduce the free energy of the transition state. Recent work from the Boxer lab has experimentally demonstrated that charges at the active site of the enzyme ketosteroid isomerase exert an electric field that contributes significantly to the catalytic effect(20). However, these charges are considered to be static. We suggest instead that the electric field at the active site of an enzyme may be modulated through the binding of charged groups at distant sites or by the presence of bioelectric fields.

Indeed, our current results indicate that this is the case. The excitation of the Ru moiety likely leads to a propagation of a polarization signal through the protein, reaching and affecting the active site. A significant effect is demonstrated here on both the binding of an antibody to the His-tag of PGK and, most remarkably, on its enzymatic activity. The effect on the activity of PGK might be due either to modulation of the binding of substrates or to an effect on the catalytic step itself- this remains to be determined. In any case, these findings point to a so-far unappreciated role of electric fields in the regulation of biological activity at the molecular level. Within the cellular environment, electric fields abound particularly near membranes, and it is possible that membrane proteins and also proteins that interact with membranes are susceptible to control mediated by charge reorganization. This discovery also suggests a novel method for generating photo-controlled enzymes and sensors, based on photoexcitation of an attached group. Currently, all proposed methods to photo-control bioactivity have relied on various conformational changes induced by photoexcitation(21, 22). Photo-controlling bioactivity through charge injection might be easier to implement. Future work will allow us to optimize the location of the photosensitizer and enhance the effect of light on activity even further and will teach us more about pathways of charge rearrangement in relation to protein function. For that purpose, we plan to identify biological systems that might be particularly susceptible to this type of activity regulation in proteins.

## Methods

**Protein expression and purification.** Yeast phosphoglycerate kinase (PGK) DNA was cloned into a pET28b vector, fused to a C-terminal 6xHis tag. For site-specific labeling of PGK, the natural cysteine (C97) was replaced by a serine. A single cysteine residue was introduced using site-directed mutagenesis, resulting in either a Q9C or a S290C PGK mutant.

Single-cysteine PGK plasmids were transformed into *E. Coli* BL21 pLysS (DE3) cells (Invitrogen), which were grown in LB media at 37 °C up to an optical density of 0.8-1. Protein expression was induced by the addition of 1mM IPTG, and cells were then incubated overnight at 25 °C. Following expression, bacteria were harvested and proteins were purified on a Ni-NTA resin (GE Healthcare), according to the manufacturer's instructions. Purified protein was dialyzed overnight in the storage buffer (20 mM sodium phosphate, 1 mM TCEP, pH 6.8) and kept at −80 °C until used.

**Protein labeling.** We took advantage of PGK's 6xHis tag, which allowed us to label the protein while bound to a Ni-NTA resin. 2 mg of histidine-tagged PGK (His-PGK) were bound to a 300 µL Ni-NTA His•Bind resin, according to the manufacturer's instructions (Millipore). The bound protein was washed with the labeling buffer (50mM Tris, pH 8) and incubated overnight with a 10-fold excess of $[Ru(2,2'\text{-bipyridine})_2(5\text{-iodoacetamido-1,10-phenanthroline})](PF_6)_2$ (Ru) at 4°C under gentle shaking. Unreacted dye was removed by washing the resin with the labeling buffer, followed by elution of labeled protein with 0.5 M imidazole in 50 mM Tris at pH 8. Finally, the eluted protein buffer was exchanged with phosphate buffered saline (PBS, Biological Industries, Reference number: 02-023-1A) using a desalting column (Sephadex G25, GE Healthcare). Protein labeling levels were determined by absorption at 450 nm. Labeled protein was kept at 4°C until used.

**Attachment of His-PGK on an Au surface.** Histidine-tagged Ru-tagged PGK (His-PGK-Ru) or unlabeled His-PGK was attached to a gold surface using dithiobis-succinimidyl propionate (DSP) as a linker. A DSP monolayer was formed on a gold surface by incubation with a solution of DSP in DMSO (4 mg/ml) for 30 min. The surfaces were rinsed with DMSO and water and were incubated into the His-PGK-Ru or unlabeled His-PGK solution (1 mg/ml) in PBS for 4 h. Then the PGK-immobilized gold surface was rinsed with PBS.

**Attachment of Histidine-tagged PGK on a glass surface.** His-PGK-Ru was attached to the glass surface of a glass-bottom Petri dish (MatTek Corporation, USA, Part No: P35G-1.0-14-C) using silane-polyethylene glycol-N-hydroxysuccinimide (silane-PEG-NHS, NANOCS, Cat. NO.: PG2-NSSL-5k) as a linker. First, a solution of silane-PEG-NHS was prepared in dry DMSO at a concentration of 1% (w/v). Glass surfaces were incubated with the linker solution for 1 h at room temperature. Then they were rinsed with DMSO, milli-Q-water and PBS, successively. 100 µl of the His-tagged PGK (1 mg/ml) solution was added to the linker-coated Petri dish and kept for 4 h. Rinsing with PBS removed the unattached protein molecules.

**Labeling of anti-His tag antibody.** In order to study the antigen-antibody reaction kinetics by observing the fluorescence of attached antibody molecules, anti-His tag antibody molecules were tagged with the dye Alexa Fluor® 647 NHS Ester (Succinimidyl Ester, ThermoFisher SCIENTIFIC, Catalog number: A20006) using the same procedure as followed in our previous paper.(11) In brief, unlabeled antibody molecules in PBS buffer were reacted with the NHS ester of the dye in a 1:1.5 ratio in presence of 0.1 M sodium bicarbonate buffer for 1 h at room temperature in the dark. Micro Bio-Spin columns with Bio-Gel P-30 (Bio-Rad) were used to remove the unlabeled dye molecules. We verified that the labeled protein did not show any optical activity at the wavelength of absorption of the Ru group using circular dichroism spectroscopy (Supporting Fig. 2).

**Interaction between His-tagged PGK and anti-His antibodies with and without illumination**

To study the antibody-antigen reaction kinetics, His-PGK-Ru modified gold surfaces were immersed in a solution of the anti-His tag antibody (0.05 µM) in PBS (pH=7.1) in a MAKTEK glass bottom petri-dish for different time intervals (2 s, 4 s, 6 s, 8 s) and immediately taken out and rinsed with buffer. The reaction was allowed to proceed either under illumination of Ru with linearly or circularly polarized light using a 470 nm laser and also without illumination. All samples were prepared twice to test reproducibility of the results. The same experiment was carried out for unlabeled His-PGK coated gold surfaces for 2 s.

To test the potential contribution of the gold surface on the antigen–antibody reaction, the above experiment was repeated using a glass surface coated with His-PGK-Ru with and without illumination.

**Microscopy measurements & data analysis.** Fluorescence imaging of the samples following reaction with antibody molecules was carried out following the same procedure used in our previous work.(11) A home-built total internal reflection fluorescence microscope (TIRFM) was used for the imaging. In each experiment, 10 different TIRFM movies were recorded on 10 different regions of 101 X 101 pixels (6.73µm X 6.73 µm). On each region, 100 ms frames were recorded until all molecules in the designated area were photo-bleached. TIRFM movies were analyzed using custom-written Matlab (MathWorks) routines. Individual spots, corresponding to individual antigen-antibody complexes, were identified in the first frame of a movie using a combination of thresholding and center-of-mass (CM) analysis as described previously(23). The intensity of the center of mass of each individual spot as a function of time was plotted, and change-point analysis was performed on to identify photobleaching steps and hence the number of emitters in each spot. Some examples are shown in Figure S3 of ref (11).

**Preparation of supported lipid bilayers for PGK activity assays.** For the preparation of supported lipid bilayers, the lipid 1,2-dioleoyl-sn-glycero-3-[(N-(5-amino-1-carboxypentyl)iminodiacetic acid)succinyl] (nickel salt) (18:1 DGS-NTA(Ni) in chloroform, 790404C, Avanti Polar Lipids Inc., USA) was aliquoted, lyophilized and then hydrated using PBS buffer at a concentration of 20 mg/ml. It was extruded through a 0.1 µm filter (Whatman Anotop, GE Healthcare, USA) to create unilamellar vesicles and stored at 4 °C. Glass-bottom Petri dishes were cleaned with 1 M NaOH (Fluka) for 40 min, and then coated with vesicle suspensions to prepare the supported bilayer(24). The lipid bilayer was incubated with 10 mM nickel chloride (Sigma) solution for 10 min, followed by attachment of His-PGK molecules to the surface.

**PGK Activity assay.** PGK molecules were adsorbed on a supported lipid bilayer for studies of enzymatic activity. This configuration facilitated continuous illumination of the molecules during the experiment. To this end, a His-PGK protein solution (1 mg/ml) was incubated over the lipid bilayer for 1 hour followed by several washes with PBS buffer to remove unattached PGK molecules. The enzymatic activity of PGK variants was measured at 25 °C by monitoring the absorbance of NADH at 340 nm using the coupled assay described by Reddy et al.(17):

$$\boldsymbol{ATP + 3\text{-}PG \xrightarrow{PGK/Mg^{2+}} ADP + 1,3\text{-}BPG}$$

$$\boldsymbol{NADH + 1,3\text{-}BPG \xrightarrow{GAPDH} GAP + Pi + NAD^{+}}$$

For the activity assay, 100 µl of a solution containing ATP (Adenosine 5′-triphosphate disodium salt hydrate, A2383, Sigma), 3-PG (D-(−)-3-Phosphoglyceric acid disodium salt, P8877, Sigma), EDTA (Ethylenediaminetetraacetic acid solution, BioPrep), NADH (β-Nicotinamide adenine dinucleotide, reduced dipotassium salt, N4505, Sigma) and GAPDH (Glyceraldehyde-3-phosphate dehydrogenase from rabbit muscle, G2267, Sigma) were added above His-PGK attached to the lipid bilayer on the glass bottom Petri dish. The reaction was initiated by adding 100 µl $MgCl_2$ solution. Final concentrations of the reagents were 5 mM ATP, 10 mM 3-PG, 1 mM EDTA, 200-600 µM NADH, 10 units/ml GAPDH and 6 mM $MgCl_2$. Aliquots were taken at specific time intervals and the decrease of NADH absorption was measured as a function of time. From the ratio of slope of the plot of NADH absorption vs. time for different reaction conditions (e.g. in absence or presence of illumination with linearly or circularly polarized light), we determined the change of enzyme activity with respect to reaction condition. The turnover number was calculated from the slope of the change of NADH concentration vs. time.

**Optical setup for excitation of sample with linear or circularly polarized light.** A 470 nm diode laser (Picoquant) was relayed through achromatic lenses to expand and collimate the beam to a diameter of 1 cm. A polarizer cube was used to modulate the polarization to excite the sample with linearly polarized light. In case of excitation with circular polarization, a quarter wave plate was introduced at the appropriate angle just before the sample chamber. We verified that at the sample light was circularly polarized to within ~10% by rotating a polarizer and measuring the power. The laser intensity (~5 $mW/cm^2$) at the sample was kept constant for linear as well as circular polarization by tuning the laser power at the source. While the relatively low laser intensity implies a low efficiency of excitation, potentially long charge recombination times are likely to lead to a significant fraction of charge-separated protein molecules.

**Acknowledgements:** This work was partially supported by a grant to G.H. from the European Research Council (ERC) under the European Union's Horizon 2020 research and innovation program (grant agreement No 742637, SMALLOSTERY), a grant from the Israel Science Foundation (no. 1250/19) and a grant from the Weizmann SABRA - Yeda-Sela – WRC program. G.H. holds the Hilda Pomeraniec Memorial Professorial Chair. R.N. acknowledges partial support from the MINERVA Foundation and from the Israel Ministry of Science and Technology. Work at Caltech was supported by the United States NIH (R01 DK019038 to H.B.G.).

**Author Contributions:** S.G. and K.B.-G. designed the project, performed all experiments and analyzed the results, D.L. prepared labeled proteins, D.S. instructed and participated in enzyme kinetics experiments, J.S. and H.B.G provided materials, G.H. and R.N. conceived and supervised the research and wrote the paper with help from all authors.

**Supporting Information.** Supporting Table 1 includes analysis of antigen-antibody binding activity assays under illumination. Supporting Table 2 includes analysis of activity assays of Ru-tagged PGK with and without. Supporting Figures 1-2 depict control experiments.

## Figure Legends

**Figure 1: A.** Structure of Phosphoglycerate kinase (3PGK): Red line represents the 6-histidine tag at the C terminus of the protein. The locations of residues 9 and 290 are depicted in raspberry and orange, respectively. **B.** Structure of Ru attached to the thiol group of a cysteine residue on the protein.

**Figure 2: Modulating PGK-antibody interaction kinetics by photoexcitation.**
**A.** Schematic of the experimental setup to study the effect of linearly and circularly polarized light on His-tagged Ru-modified PGK- anti-His antibody interaction kinetics. **B-E.** Fluorescent images of individual complexes formed between His-tagged PGK molecules labeled with Ru at residue 9 and adsorbed on a gold surface and alexa-647 labeled anti-His antibodies in presence and absence of illumination with linearly polarized light for 2 s and 8s. **F.** Kinetics of PGK-antibody association with (red) and without (blue) illumination, as obtained by counting molecules in fluorescent images. **G.** No effect of illumination on PGK-antibody association kinetics was observed in the absence of Ru. **H.** The experiment of panels B-E was repeated with PGK adsorbed on glass, with similar results. **I.** Effect of the polarization of the light on the photoinduced enhancement of PGK-antibody association kinetics. RCP- right circular polarization, LCP- left circular polarization, LP- linear polarization. **J.** Only a minor illumination effect was observed when Ru was moved to residue 290. In G-J molecules were counted 2 s following the initiation of the reaction. At least 9 regions were counted in each sample. Experiments were repeated three times (see Supporting Table 1 for all values). Error bars represents standard errors of mean.

**Figure 3: Modulating enzymatic kinetics by illumination.**
**A.** Schematic of the experimental setup to study the effect of illumination on enzymatic kinetics of Ru-modified PGK. PGK molecules were attached to a lipid layer supported on a glass surface through their His-tags. The enzymatic reaction of PGK is depicted in the cartoon. Enzymatic activity was measured at 25 °C using a coupled assay (see Methods) and the absorbance of NADH at 340 nm was monitored. **B.** A strong reduction in enzyme kinetics was observed upon either LP or LCP illumination of PGK modified with Ru at position 290, but not under RCP illumination, as compared to no light (NL). **C.** The slope of reaction progression changed when the initial LP illumination was stopped after 5 min (full symbols). In the absence of Ru, no effect of light was observed (empty symbols). **D.** The effect of light was smaller when Ru was at position 9. As in B, the effect was observed under LP or LCP illumination, but not under RCP illumination. **E.** As in C, but with Ru at position 9. Only the linear regions of the activity curves were fitted. Experiments were repeated three times, and this figure shows only one set. For values obtained from all experimental sets, see Supporting Table 2.

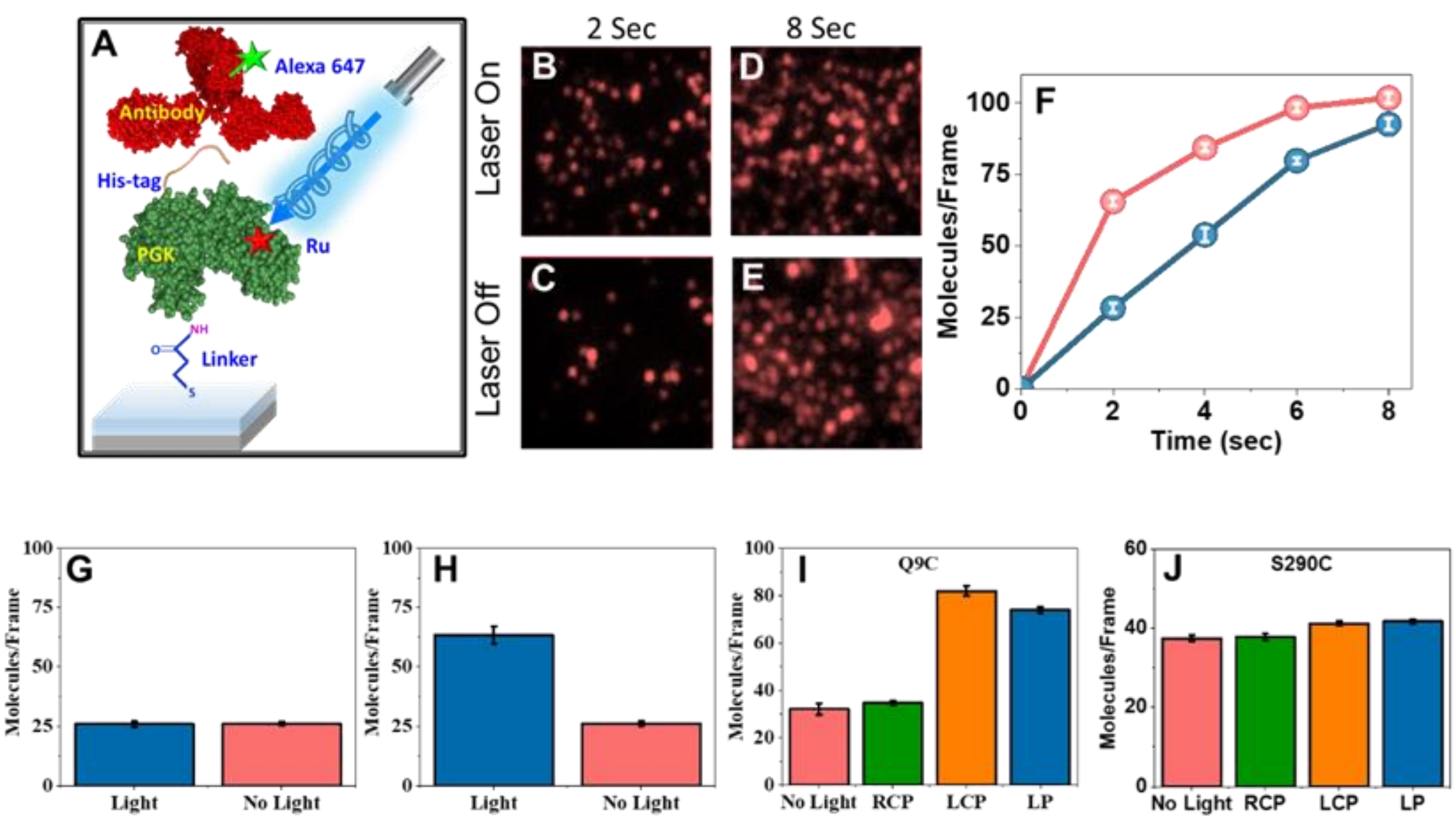

**Figure 2: Modulating PGK-antibody interaction kinetics by photoexcitation.**
**A.** Schematic of the experimental setup to study the effect of linearly and circularly polarized light on His-tagged Ru-modified PGK- anti-His antibody interaction kinetics. **B-E.** Fluorescent images of individual complexes formed between His-tagged PGK molecules labeled with Ru at residue 9 and adsorbed on a gold surface and alexa-647 labeled anti-His antibodies in presence and absence of illumination with linearly polarized light for 2 s and 8s. **F.** Kinetics of PGK-antibody association with (red) and without (blue) illumination, as obtained by counting molecules in fluorescent images. **G.** No effect of illumination on PGK-antibody association kinetics was observed in the absence of Ru. **H.** The experiment of panels B-E was repeated with PGK adsorbed on glass, with similar results. **I.** Effect of the polarization of the light on the photoinduced enhancement of PGK-antibody association kinetics. RCP- right circular polarization, LCP- left circular polarization, LP- linear polarization. **J.** Only a minor illumination effect was observed when Ru was moved to residue 290. In G-J molecules were counted 2 s following the initiation of the reaction. At least 9 regions were counted in each sample. Experiments were repeated three times (see Supporting Table 1 for all values). Error bars represents standard errors of mean.